\documentclass[reprint,superscriptaddress,amsmath,amssymb,aps,prb]{revtex4-2}

\usepackage{graphicx}
\usepackage{dcolumn}
\usepackage{bm}
\usepackage{color}
\usepackage{ulem}

\def\vec#1{\bm{#1}}
\def\ket#1{|{#1}\rangle}

\begin{document}

\title{Terahertz response of confined electron-hole pair: crossover between strong and weak confinement}%
\author{Filip Klimovi\v{c}}%
\email{filip.klimovic@matfyz.cuni.cz}%
\affiliation{%
 Faculty of Mathematics and Physics, Charles University, Ke Karlovu 3, CZ-12116 Prague 2, Czech Republic
}%
\author{Jens Paaske}%
\affiliation{%
 Niels Bohr Institute, University of Copenhagen, Universitetsparken 5, DK-2100 Copenhagen Ø, Denmark
}%
\author{Tom\'{a}\v{s} Ostatnick\'{y}}%
\affiliation{%
 Faculty of Mathematics and Physics, Charles University, Ke Karlovu 3, CZ-12116 Prague 2, Czech Republic
}%
\date{\today}

\begin{abstract}
    We analyze theoretically THz response of an electron-hole pair confined in a semiconductor nanoparticle.
    We show that the interplay of particle confinement and electron-hole Coulomb interaction leads to
    significant renormalizations and energy shifts in THz linear conductivity of the nanocrystal.
    We develop and evaluate models in the strong and the weak confinement regime in order to correctly
    address the effect of Coulomb interaction. In the weak confinement regime, we find solutions of the
    problem in a form similar to the Wannier wavefunction whose spatial extent is reduced as a consequence 
    of the confinement. The resulting states are scalable down to the strong confinement regime, enabling
    a theoretical description of the exciton response for arbitrarily sized nanoparticles.
\end{abstract}

\maketitle


\section{\label{sec:introduction}Introduction}

Nanostructured materials nowadays play an important role in many technologies, in particular in electronics.
Knowledge and ability to predict their properties makes it possible to design advanced composites with
the nanostructures as the basic building blocks on the one hand, and to develop experimental methods for
their characterization on the other. Contemporary technologies have unlocked the terahertz spectral region 
for effective THz spectroscopical methods and also for newly developed fast electronic devices by 
increasing their bandwidth. The ability of efficient modeling of the response of confined electronic 
systems to THz fields becomes therefore essential for both the material research and spectroscopy.

A special focus was put on the single-particle response of a semiconductor nanocrystal (see reviews 
\cite{lloyd-hughes2012,kuzel2019}). Besides a rich variety of effective semi-classical models,
microscopic theories were also developed, starting with a classical particle-in-a-box problem 
\cite{nemec2009}, which was followed by quantum-mechanical models 
\cite{ostatnicky2018,ostatnicky2019,quick2023}. The applicability of the quantum models to 
experimental data has been demonstrated and a direct proof of the appearance of a sharp resonance 
due to quantization of the confined electron energy has been shown 
\cite{pushkarev2017,pushkarev2022}. Within the description beyond the single-particle picture, it is 
widely believed that the electron kinetic energy caused by the quantum confinement exceeds the 
Coulomb interaction energy \cite{banyai1993} in small nanocrystals and therefore two-particle states 
(in particular excitons) can be treated as mostly separate electrons and holes with a mild energy 
correction due to the electrostatic interaction (mean field approximation). Studies beyond the mean
field approximation accounting for the finite confining potential \cite{einevoll1992,iotti1997,david2022}
or parabolic potential \cite{que1992,holtkemper2020} exist, including realistic microscopic calculations of exact
exciton energies within many-band models. Coulomb-induced renormalizations in THz response spectra
were not, however, discussed so far.

Our particular interest in exciton THz response in semiconductor nanocrystals stems from the fact that 
they are the basic electronic excitations of intrinsic semiconductor quantum dots in optical pump -- 
THz probe experiments. Coincidentally, inter-subband transitions between exciton excited states lie 
in the THz spectral range \cite{kaindl2003,li2020,pushkarev2022,leinss2008,steinleitner2017} and so 
the interplay of the electron quantization and Coulomb electron-hole (e-h) interaction favors THz 
experiments as a suitable experimental tool to probe the exciton internal structure. Besides the 
spectroscopic point of view, excitons are also interesting for their potential applications 
\cite{anantharaman2021,efros2021}. They were shown, for example, to reveal long-distance propagation 
in low-dimensional structures \cite{tagarelli2023,sanderson2019} and quantum dots can serve as 
excitonic qubits \cite{harankahage2021}. 

It is reasonable to define three different regimes for the confined e-h pair, depending on the ratio between
the kinetic and electrostatic interaction energy. There are several approaches that span from
the usage of Bohr radii of individual particles \cite{efros1982} to the definition based on the
comparison of the confinement radius $A$ and exciton Bohr radius $a_X=4\pi\hbar^2\varepsilon/e^2\mu$ 
\cite{richter2017} where $\varepsilon$ denotes the material background permittivity, $e$ is the 
elementary charge, $\mu=m_em_h/(m_e+m_h)$ is the reduced e-h mass and $m_{e,h}$ are individual 
effective masses of an electron/hole. Our definition of the three regimes is based on the exciton 
Bohr radius $a_X$ since we trace qualitative changes to the exciton wavefunction rather than renormalizations
of the state energy, as will be discussed later. We thus define:

1. {\sl weak confinement regime (WCR)} for $a_X \ll A$, where it is justified to separate the center-of-mass 
(COM) and relative motion \cite{dresselhaus1956}. The bound states of the relative motion are described by hydrogen-like solutions (Wannier exciton), whereas the confined COM motion quantization 
introduces a small correction to the energy levels. The COM subbands 
have been observed in photoluminescence experiments \cite{lage1991,lee2018}. Lage {\it et al.} also 
suggest \cite{lage1991} the need to consider the finite size of the exciton when comparing the 
COM quantization length extracted from the experimental data to the geometric size of the 
nanostructure. The question of the finite size of the exciton becomes even more apparent when 
considering so-called Rydberg excitons \cite{chernikov2014,kazimierczuk2014,orfanakis2022} and
their confinement \cite{orfanakis2021}, keeping in mind that their Bohr radius
can scale up to units of microns.

2. {\sl strong confinement regime (SCR)} in the case of $A<a_X$, where the kinetic energy of the confined 
electron and hole dominates and the Coulomb interaction is usually treated by a perturbative approach 
\cite{banyai1993}. Photoluminescence experiments show an increase of the exciton energy 
\cite{rossi2020,haldar2017} due to the confinement. Concerning the THz response, 
recent quantum-mechanical models \cite{ostatnicky2018,ostatnicky2019, 
quick2023} treat the individual charge carriers as independent, neglecting the Coulomb 
interaction completely.

3. {\sl intermediate confinement regime (ICR)} for $a_X\lesssim A$, in which case a particle (electron or hole)
with lesser effective mass (usually electron) is delocalized and effectively creates a 
potential for the heavier particle. The latter is then localized near the center of the 
quantum dot. It is usually treated approximately as a charged impurity-like system 
\cite{chuu1992,bellessa1999,ziemkiewicz2020}, further reducing the complexity 
of the problem. Alternately, full separation of relative and COM motion can be introduced,
regarding the exciton as a hard ball of finite size confined inside a potential of finite dimensions.
The relative coordinate is then subject to a Wannier-like equation and the COM motion
is found in the form of a particle-in-a-box within the crystal confining potential,
reduced by a so-called dead layer \cite{fonoberov2004,raikh2021}, given by the size of the hard ball.

In this paper, we focus on the details of e-h correlations caused by their inherent 
Coulomb interaction beyond the mean field approximation. We develop a theory which explains and
predicts how these correlations display in THz linear response spectra. Unlike the exciton linear
response at optical frequencies, which is capable of probing the overall e-h pair energy and eventually the
oscillator strength for interband electron transitions, THz detection has direct access to individual
electron-/hole-like resonances. To
this end, we find a solution of the two-particle Schr\"{o}dinger equation beyond the mean field 
approximation and calculate the THz linear response spectra. We show that substantial 
renormalizations of the resonance energies and oscillator strengths appear in the SCR.
We find a smooth transition of a shrunken exciton wavefunction from the SCR to the WCR,
which can also be useful for describing spatially extended Rydberg states
\cite{konzelmann2020,orfanakis2021}. Finally, we discuss an apparent contradiction between the 
WCR model and the SCR model in the small crystal limit: the former one ascribes a nonzero oscillator
strength only to the electron-hole relative motion, while the COM is decoupled from
the electromagnetic field due to the exciton charge neutrality. For this reason, a series of resonances 
occurring from only internal degrees of freedom is naively expected in exciton response spectra in 
nanostructures \cite{quick2022}. On the other hand, SCR provides a separate and almost
independent motion of both electrons and holes, providing two series of resonances. We
show that these two models can be smoothly connected in the ICR and discuss the physical interpretation.

\section{\label{sec:model}Model}

We consider a single spinless e-h pair confined in a quantum dot. This situation resembles the
response of an optically inactive ground state of an electron and a heavy-hole in
intrinsic GaAs. Note that our conclusions below are valid also for optically active states
where the exchange interaction introduces mild corrections to the
quantitative results. We employ the two-band effective mass (envelope function) approximation in
which our model system can be represented in general by the Hamiltonian:
\begin{multline}
    H=\frac{\hbar^2}{2m_e}\nabla_e^2 -\frac{\hbar^2}{2m_h}\nabla_h^2 +V(\bm{r}_e) +V(\bm{r}_h)-\\
    -\frac{e^2}{4\pi \varepsilon\left| \bm{r}_e-\bm{r}_h \right|}\,,\label{eq:ham_gen}
\end{multline}
where the first two terms on the right-hand side represent the kinetic energy operators for the 
electron and the hole, respectively, the next two terms are their respective confinement potentials, 
and the rightmost term stands for their mutual Coulomb interaction energy. The vectors $\bm{r}_e$ and 
$\bm{r}_h$ denote the electron/hole respective positions. In the present work, we assume spherical 
geometry and infinite rectangular profile of the confinement potential:
\begin{equation}
    V(\bm{r}_{e,h})= V(r_{e,h})=\begin{cases}
        0 \quad &\text{for } r_{e,h} < A\\
        \infty \quad &\text{for } r_{e,h} \ge A
    \end{cases}
    \label{eq:conf_potential}
\end{equation}
where $A$ is the radius of the confinement potential.
The electron coherence length is assumed to be large compared to the confinement size $A$ and 
therefore a description in terms of a full two-particle wavefunction is required.
Solving for the eigensystem of the general Hamiltonian \eqref{eq:ham_gen} is a formidable task and requires
different strategies for different regimes.

In order to account for an eventual permittivity contrast between the nanocrystal ($\varepsilon_1$) and the surrounding material ($\varepsilon_2$), it is possible to include additional terms of the Coulomb potential in Eq.~(\ref{eq:ham_gen}) in a form of an infinite series, according to chapter 2-5 of Ref. \onlinecite{banyai1993} and follow the procedure described below with the full expression. As the first approximation, permittivity mismatch causes an effective decrease/increase of the crystal permittivity $\varepsilon$ if the crystal permittivity is higher/lower than that of the surrounding material, thus modifying the electron-hole interaction energy. The asymptotic scaling of the interaction energy correction is $\propto -e^2(\varepsilon_1 / \varepsilon_2-1)/\varepsilon_1 A$, and therefore its contribution is negligible in the WCR compared to the major interaction term $\sim-e^2/4\pi\varepsilon_1a_X$, while it may be significant in the SCR.

\subsection{\label{subsec:strong_conf}Strong Confinement}

In nanocrystal quantum dots that are small compared to the exciton Bohr radius,
the most important energy scale comes from the bare single particle-in-a-box energies. These
are addressed by the Hamiltonian:
\begin{equation}
  H_j^{(0)}=-\frac{\hbar^2}{2m_{j}}\nabla_{j}^2+V(r_{j})\,,
\end{equation}
where the index $j$ denotes an electron or hole. The solutions appear to be:
\begin{eqnarray}
  E_{jnl}^{(0)} &=& \frac{\hbar^2k_{nl}^2}{2m_j}\,,\label{eq:bareEnergy}\\
    \psi_{jnlm}(r,\theta,\phi) &=& \mathcal{N}_{nl} j_l(k_{nl}r) Y_{l}^m(\theta,\phi)\,,
    \label{eq:particle_in_a_box_wf}
\end{eqnarray}
with $n=1,2,3,...$; $l=0,1,2,...$ and $m=-l,...,l$. $\mathcal{N}_{nl}$ is the normalization 
constant, $j_l$ are the spherical Bessel functions describing the radial part of the wavefunction 
and $Y_{l}^m$ are the spherical harmonics for the angular part. The wavevectors $k_{nl}$ satisfy the 
boundary condition $j_l(k_{nl}A)=0$. In the following, we will label the single-particle states
for example $1p_z$, which means $n=1$, $l=1$ and $m=0$, where the letters are used in order
to describe the usual orbital symmetry.
The bare two-particle states are constructed as direct products
$\left|e,h\right\rangle = \left|e\right\rangle \otimes \left|h\right\rangle$
and the total kinetic energy of the two-particle state is the sum of the two single-particle energies.

The exciton states, which are the eigenstates of the Hamiltonian (\ref{eq:ham_gen}), are then expanded
in terms of the bare two-particle states:
\begin{equation}
    \left|\alpha\right\rangle = \sum_i c_i \left| e, h \right\rangle_i\,.
    \label{eq:eigenstates}
\end{equation}
Here, the index $i$ stands for a combination of three electron plus three hole quantum numbers.
The appropriate expansion coefficients $c_i$ are found by diagonalization of the Hamiltonian (\ref{eq:ham_gen}) within the basis of bare two-particle states. To gain the full solution of the problem, a full set of all basis states would be necessary. We will call this hypothetical situation an {\sl exact model} in the following. In a real calculation, however, it is only possible to use a finite number of states, further restricted by the demands on computational power. For that reason, one might select only several relevant two-particle states in the expansion, which we will call the {\sl SCR model}. Contrary to the exact model, its validity will be limited to sufficiently small crystals where the Coulomb interaction energy is small with respect to level splitting due to the confinement.

\subsection{\label{subsec:weak_conf}Weak Confinement}

The WCR model is based on the Wannier bulk solutions of the electron-hole problem.
Relative and COM motion are separated by the standard transformation of coordinates:
\begin{eqnarray}
    \bm{r} &=& \bm{r}_e - \bm{r}_h\,,\\
    \bm{R} &=& \frac{m_e \bm{r}_e + m_h \bm{r}_h}{m_e+m_h}\,,
\end{eqnarray}
leading to hydrogen-like wavefunctions for the relative part with bound state energies $E_n$ and Bohr
radii $a_n$ in the absence of any confining potential:
\begin{eqnarray}
    E_n &=&-\frac{\mu e^4}{32 \pi^2 \varepsilon^2 \hbar^2} \frac{1}{n^2}\,, \\
    a_n &=& \frac{4 \pi \varepsilon \hbar^2}{\mu e^2} n^2\,,
    \label{eq:bohr_radius}
\end{eqnarray}
where $\mu$ is the reduced effective mass of the e-h pair. We denote the eigenstates similar
to the general hydrogen-like atom problem as $(1s)_w$, $(2p_x)_w$ etc. We introduced the label $(\ldots)_w$
in order to distinguish between WCR and SCR notation. While, for example, the first excited state in 
the SCR is $1p$, it is $(2p)_w$ in the WCR. In the presence of the confinement potential, such {\sl 
exact} separation is not possible due to the broken translational symmetry. Under the coordinate 
transformation, the general Hamiltonian \eqref{eq:ham_gen} is:
\begin{multline}
    H=
    -\frac{\hbar^2}{2M}\nabla_R^2 -\frac{\hbar^2}{2\mu}\nabla_r^2  -\frac{e^2}{4\pi \varepsilon r}+\\
    +V\left(\bm{R}+\frac{\mu}{m_e}\bm{r}\right) +V\left(\bm{R}-\frac{\mu}{m_h}\bm{r}\right)\,,
    \label{eq:ham_weak_conf_before_sep}
\end{multline}
where $M=m_e+m_h$ is the total mass of the e-h pair. We perform an {\sl approximate} 
separation by assuming that the exciton is a solid ball of a certain radius $\varrho$. The 
confinement potential for the exciton then becomes $V(R+\varrho)$. It is then usually safe to set
$\varrho$ to a fixed value within the WCR model, for example to the exciton Bohr radius.
Considering the Rydberg states whose radius scales as the principal quantum number squared, we find that
the value of $\varrho$ exceeds the crystal size at some point. We therefore have to consider an adaptive 
$\varrho$ that then turns out to be rather an operator. We set the radius $\varrho$ value to
the greater of the distances of the electron and hole from their COM:
\begin{equation}
    \varrho(r)=\max\left( \frac{\mu}{m_e}r;\frac{\mu}{m_h}r \right)
    \label{eq:radius_rho}
\end{equation}
and the approximate Hamiltonian then becomes:
\begin{equation}
    H=
    -\frac{\hbar^2}{2M}\nabla_R^2 +V\big(R+\varrho(r)\big) -\frac{\hbar^2}{2\mu}\nabla_r^2
    -\frac{e^2}{4\pi \varepsilon r}\,.
    \label{eq:ham_weak_conf}
\end{equation}
The solution to the Schr\"{o}dinger equation is derived in the Appendix~\ref{app:a}.
Part of the separation procedure is the assumption that the relative and COM motions are 
uncorrelated and that the COM part then becomes a particle-in-a-box problem with the 
variable $r$ as a parameter affecting only the confinement volume. The COM ground-state 
kinetic energy \eqref{eq:com_ground_state_app} is
\begin{equation}
    \chi(r) = \frac{\hbar^2}{2M}\left(\frac{\pi}{A-\varrho(r)}\right)^2\,.
\end{equation}
Apparently, the larger the extent of exciton $\varrho(r)$, the higher the kinetic energy of the 
COM motion $\chi(r)$ (and in turn the total energy of the system), which thus effectively 
creates an additional potential term for the relative motion. The angular part of the relative 
motion remains untouched, therefore one can anticipate a separated relative motion wavefunction 
$\psi(\bm{r})$ in the form
\begin{equation}
  \psi_{nlm}(r,\theta,\phi) = \mathcal{N}_{nl}\mathcal{R}_{nl}(r) Y_{l}^m(\theta,\phi)\,,
  \label{wavefunc}
\end{equation}
where $Y_{l}^m$ are the spherical harmonics. The radial part of the Schr\"{o}dinger equation 
\eqref{eq:radial_eq_app} for $\mathcal{R}$ and the total energy $E$ remains to be solved numerically:
\begin{multline}
  \left[ -\frac{\hbar^2}{2\mu}\frac{\partial^2}{\partial r^2} 
  -\frac{\hbar^2}{2\mu}\frac{2}{r}\frac{\partial}{\partial r} + \frac{\hbar^2}{2\mu}\frac{l(l+1)}{r^2}
  \right.-\\
  \left.-\frac{1}{4\pi \varepsilon}\frac{e^2}{r} + \chi(r) \right]\mathcal{R}(r) = E\; \mathcal{R}(r)\,.
\end{multline}

\subsection{\label{subsec:thz_response}THz Response}

We evaluate the exciton linear response to the external electric ac field in terms of the (isotropic)
ac conductivity \cite{pushkarev2017,ostatnicky2018}, which we evaluate as its $\sigma_{xx}$ component:
\begin{equation}
    \sigma(\omega) = {-}\frac{i \omega e^2}{\mathcal{V}} \sum_{k,l} \frac{f_{kl}}{\hbar}\frac{\left| 
\left\langle k|d_x|l \right\rangle \right|^2}{\omega-\omega_{kl}+i\gamma}\,,
    \label{eq:conductivity}
\end{equation}
where $\mathcal{V}$ is the nanocrystal volume and $\gamma$ is the electron momentum scattering rate. 
$\left\langle k|d_x|l \right\rangle$ is the dipole moment between the exciton states given by 
multi-indices $k,l$. The dipole moment operator is $\bm{d} = -e(\bm{r}_e-\bm{r}_h)$, 
$f_{kl} = f_k-f_l$ stands for the difference in occupations of the two states, and $\hbar\omega_{kl} 
= E_k-E_l$ is the energy of the dipole transition. We further normalize the conductivity in order to
quantify the response of a single e-h pair: $\bar{\sigma}=\sigma\mathcal{V}/2e$.

Note that conductivity $\sigma$ refers to the ac {\it displacement} currents as a 
response to the external THz field. The net electrical current density of the transport of entire 
excitons is zero and, therefore, it is not coupled to the external field.

\section{\label{sec:results}Results}

\subsection{Conductivity spectra}

\begin{figure}
    \includegraphics[width=1.\linewidth]{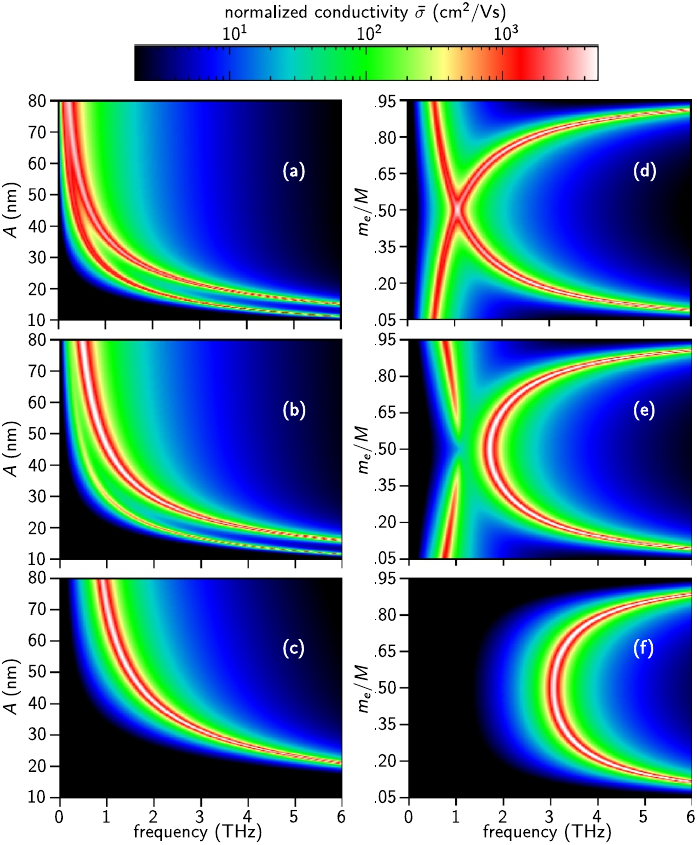}%
    \caption{Spectra (real part) of the normalized linear THz conductivity of a single e-h pair   in a spherical nanoparticle in the ground state, employing different approximations: (a),~(d) bare non-interacting electron-hole pairs; (b),~(e) strong confinement model; and (c),~(f) weak confinement model. First column displays dependency of the spectra on nanocrystal  radius $A$ with $m_e=0.35M$ ($a_X=15$~nm), second column shows dependency of the spectra on electron effective mass, keeping $m_e+m_h=0.2m_0$ and setting crystal radius $A=30$~nm.}
    \label{fig:spc}
\end{figure}

We show in Fig.~\ref{fig:spc} the real part of normalized conductivity spectra $\bar{\sigma}(\omega/2\pi)$ calculated using Eq.~\ref{eq:conductivity}. The material parameters were chosen to best highlight differences between the models used and do not resemble any particular material. The results for real materials are presented below in Fig.~\ref{fig:lines}. For the purpose of Fig.~\ref{fig:spc}, we set the exciton mass $M=0.2m_0$ where $m_0$ is the free electron mass. Particle masses were chosen $m_e=0.35M$ and $m_h=0.65M$, the relative permittivity was $\varepsilon=12.85$ (a typical value for GaAs and similar to other semiconductors) so that the exciton Bohr radius is $a_X=15$~nm, the relaxation rate $\gamma=1/2.7\ \mathrm{ps}^{-1}$ and the temperature was set to $T=0$~K so that only the ground state is populated. We also restricted the number of states in the calculation to a minimal basis of single-particle states $1s$ and $1p_{x,y,z}$, which is capable of describing the major resonances.

The dependence of hypothetical conductivity spectra of non-interacting electron and hole on the crystal radius is displayed in Fig.~\ref{fig:spc}(a). We clearly resolve two distinct resonances which do not merge even at the crystal size that is ascribed to the WCR. The WCR model, whose results are plotted in Fig.~\ref{fig:spc}(c), reveals only a single $(1\mathrm{s})_w\to(2\mathrm{p_x})_w$ resonance in the whole range of crystal sizes. Finally, the SCR model in Fig.~\ref{fig:spc}(b) correctly reveals a single resonance for large crystals, while there are two separate resonances in small crystals.

To address the interplay of the electron, hole and exciton internal transitions in more detail, we plot spectra for noninteracting e-h pair, the SCR and WCR model, respectively, in Fig.~\ref{fig:spc}(d), (e), (f), keeping fixed nanocrystal radius $A=30$~nm and varying the electron mass. For the non-interacting electron and hole in Fig.~\ref{fig:spc}(d), each of the particles gives rise to one resonance whose spectral position continuously shifts towards higher frequency as the mass of the particle decreases. In the symmetric case $m_e=m_h$, the individual resonances overlap, giving rise to the crossing. In the WCR model in Fig.~\ref{fig:spc}(f), on the contrary, the Wannier-like exciton state reveals a single resonance for any ratio of the particle masses because the COM degree of freedom of the two-particle state is decoupled from electromagnetic field due to exciton charge neutrality. The strong confinement model in Fig.~\ref{fig:spc}(e) finally includes the inter-particle Coulomb interaction while fully accounting for coupling of all degrees of freedom to the external field. It results in distinct electron- and hole-like resonances in the cases $m_e/m_h\gg1$ or vice versa. Proceeding towards electron-hole degeneracy, the lower-frequency resonance ceases. At the same time, the high-frequncy resonance position apparently shifts to a higher frequency compared to the non-interacting case. It does not reach the frequency observed in the WCR model because neither of the models is exact (the SCR model due to the incomplete set of basis states).

\begin{figure}
    \includegraphics[width=1.\linewidth]{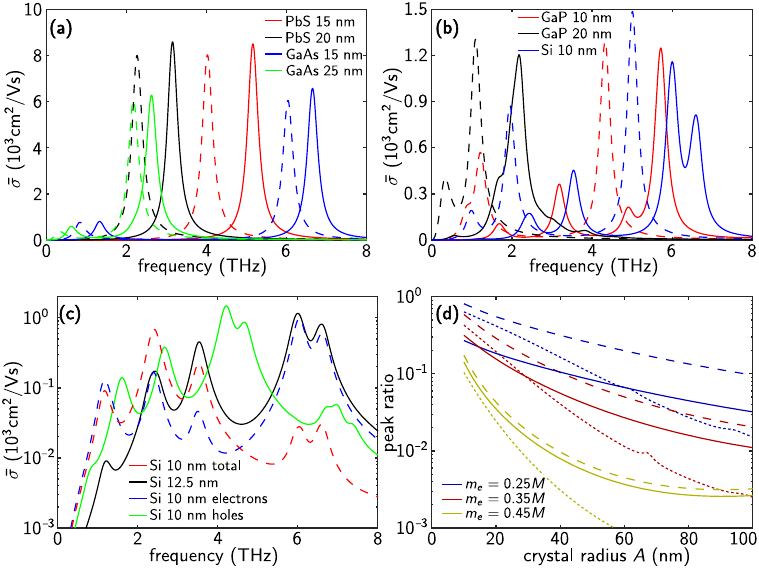}
    \caption{The real part of normalized conductivity spectra of non-interacting electron-hole pairs (dashed lines) and result of strong-confinement model (solid lines) for various direct-gap nanoparticles (a) and indirect-gap nanoparticles (b). (c) Spectra for 10~nm (black) and 12.5~nm (green) Si nanoparticles and separate contributions of the electron (blue) and hole (red) to the response function for the 10~nm particle. (d) Ratio of peak values of two resonances observed in Fig.~\ref{fig:spc}(b) (solid lines) and the same ratio multiplied by $m_h/m_e$ (dashed lines). Dotted lines display the same ratio as the dashed lines, using an extended basis of 729 states. The exciton mass is $M=0.2m_0$ and the electron mass $m_e$ is a parameter.}
    \label{fig:lines}
\end{figure}

Apart from the general plots in Fig.~\ref{fig:spc}, we give examples of normalized conductivities (real part) of real materials with a fixed ratio $m_e/M$ in Fig.~\ref{fig:lines}(a) and (b). For example, this ratio reaches the value 0.13 in GaAs, in which case the Coulomb-induced renormalizations are not very prominent in nanoparticles. In PbS, on the other hand, the masses $m_e=m_h=0.5m_0$ indicate a strong effect of the Coulomb interaction on the e-h dipole moment and also on the spectral position of the ac conductivity resonance even in strongly confined systems, see Fig.~\ref{fig:lines}(a). Coulomb-induced mixing of states in indirect gap semiconductors is even more complicated due to the anisotropic electron mass where one value usually exceeds the hole effective mass and the other one is lower. Examples of spectra of normalized conductivities for Si and GaP are shown in Fig.~\ref{fig:lines}(b). All spectra of indirect semiconductors were modeled using the SCR model, assuming a cubic shape of the crystal in order to simplify an implementation of the symmetry breaking by the anisotropic electron effective mass. The crystal edge length was set to $A'=1.7A$ where $A$ is the radius of an equivalent spherical particle. This ratio was chosen in order to best capture the spectral position of the transition between the ground and the first excited state of an electron-hole pair.

As already seen in Figs.~\ref{fig:spc}(b) and (e), electrostatic e-h correlations lead to shifts
of the resonant frequencies and to the redistribution of their oscillator strengths. The ratio of peak values in the conductivity spectra defined as the lower-frequency peak value over the higher-frequency peak value can then be used as a measure of the Coulomb-induced mixing. We plot this ratio as a function of nanocrystal radius for exciton mass $M=0.2m_0$ (the same as in Fig.~\ref{fig:spc}) and different values of relative electron mass $m_e/M$ in Fig.~\ref{fig:lines}(d) by solid lines. Clearly, this ratio decreases as the crystal size increases, indicating a clear transition from the case of mostly independent electron and hole to a bound exciton-like state with only one resonance related to the relative e-h motion. It can be shown that single-electron conductivity is inversely proportional to the electron effective mass \cite{ostatnicky2019} and therefore it is reasonable to multiply the ratio in Fig.~\ref{fig:lines}(d) by a factor $m_h/m_e$ (assuming $m_h>m_e$) as indicated by dashed lines. For zero nanocrystal size, i.e. when quantum confinement completely overrides the Coulomb interaction, the lines approach unity as expected. Their decrease with the nanocrystal radius is more pronounced for $m_e\approx m_h$. In the exact equality of the particle masses, the lower-frequency resonance should cease for any crystal size, in accord with Fig~\ref{fig:spc}(e).

\subsection{Intermediate confinement}

The two-particle wavefunction cannot be found analytically in the ICR between the SCR
and the WCR. Numerical diagonalization of the Hamiltonian can be employed but
it is demanding on the computational resources and does not give any general conclusions.
The shrunken Wannier-like solution defined in Eq.~(\ref{wavefunc}) can serve as a good
candidate for at least an estimate of the spectra in the
region of ICR, not covered by the SCR and WCR models. We find in Fig.~\ref{fig:lines}(d) that the height of the minor conductivity peak
is less than 20\% of the major one at crystal radii equal to the appropriate exciton Bohr radius
which equals 18, 15 and 14~nm for $m_e/M=0.25$, 0.35 and 0.45, respectively. The WCR model therefore
does not accurately describe the whole spectrum, but we still believe that it might be used for
an approximate description of the major conductivity peak.

\begin{figure}
    \includegraphics[width=1.\linewidth]{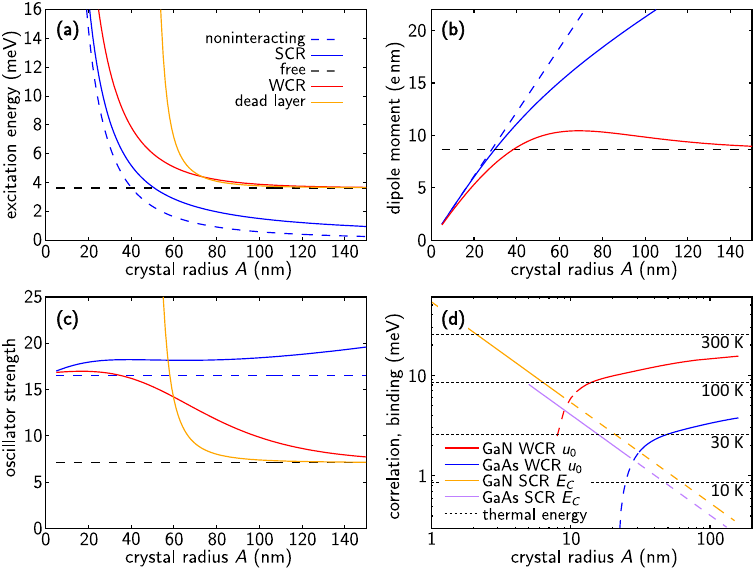}
    \caption{(a), (b), (c) comparison of observables evaluated using the strong (blue lines), weak (red line),
    dead layer (orange) and free exciton model (black line). Dotted and solid blue lines
    denote the model without and with Coulomb interaction included, using the minimal basis. Material parameters are those of bulk GaAs.
    The displayed observables are energy of the considered dipole transition (a), its dipole moment (b)
    and the oscillator strength (c).
    (d) correlation $|E_C|$ and binding energy $|u_0|$ compared to thermal energy $k_B 
    T$. The dashed lines represent regimes in which the applicability of the given model can be 
    questioned in favour of the other. In panel (d), both axes are in logarithmic scale.}
    \label{fig:mmt}
\end{figure}

To demonstrate suitability of the WCR model in ICR, 
we show in Fig.~\ref{fig:mmt}(a), (b) and (c) a smooth transition of relevant observables as provided by our WCR
model (red line), considering a GaAs nanocrystal.
Outputs of the model are compared to those of the SCR model (blue lines) (relevant in the small crystal limit) 
on one side and a free exciton (black line) applicable
for an infinite crystal on the other. The blue lines are shown for the case with (solid) and without
(dotted) Coulomb interaction included, using the minimal basis.
We also added the result of the dead layer model \cite{fonoberov2004,raikh2021} using orange line.
The WCR model smoothly connects to the limiting cases on both sides. Furthermore, it 
correctly finds a balance between shrinking the exciton and reducing the effective extent of
confining potential for the COM motion so that it avoids the energy divergency of the
dead layer model (which considers exciton of a fixed size) as seen in Fig.~\ref{fig:mmt}(a), (b) and (c), orange line.
Our approach to the weak confinement model thus serves as a suitable candidate for the description of
confined excitons in general, including the spatially extensive Rydberg states.

\section{\label{sec:discussion}Discussion}

\subsection{Strong confinement}

We can observe in Fig.~\ref{fig:spc}(b) that the strong confinement model predicts a smooth transition from two main peaks in the response of a small crystal occupied by an e-h pair to a single-peaked conductivity spectrum of a crystal large compared to the exciton Bohr radius. Coulomb interaction within the electron-hole system not only blueshifts the spectral position of resonances, but also renormalizes their oscillator strengths. The shift of the resonant frequencies is not much apparent in a small crystal, comparing the plots in Fig.~\ref{fig:spc}(a) and (b) but it becomes significant above crystal radius 20~nm. The blueshift from 0.22~THz for noninteracting particles to 0.45~THz in the SCR model can be resolved for crystal radius 80~nm. Even the SCR model underestimates the resonant frequency, as illustrated in Fig.~\ref{fig:spc}(c) where we show predictions of the WCR model. This model should be more relevant for such a large crystal radius and reveals the resonance at the frequency 1~THz. This discrepancy between the SCR and WCR models is caused by the limited set of bare states considered in evaluation of the SCR model: it can be verified numerically that an increase of the number of basis states leads to the shift of the resonance towards the result of the WCR model. As noted above, suitability of such an approach is questionable for large crystals. An exact criterion for a terminal crystal radius is hardly possible to derive; however, we may argue that the Coulomb interaction overrides the electron and hole mostly independent motion once the particles start orbiting in a state similar to the Wannier exciton. We thus find an upper limit for the crystal radius in the SCR model as the exciton Bohr radius $a_X$. Our numerical calculations (not shown here) further verified that this estimate provides a reasonable criterion.

The renormalizations of the dipole moments ascribed to the particular resonances discussed above show a striking influence of the crystal dimension on the THz linear response of the e-h pair. In optical spectra, for example, it is mainly the shift in the energy of the dipole transition (due to the quantum confinement and Coulomb interaction) that is of importance, not the change in the dipole moment itself. The band-to-band electron transition rates are not very 
sensitive to Coulomb mixing \cite{banyai1993}. In the THz spectral region, on the other hand, we probe separate electron and hole responses, which then interfere and hence can add up or subtract, depending on the particular mixing of the states due to the Coulomb interaction. The dipole moments of particular resonances in the THz conductivity spectra thus carry important information about the system size and geometry, and we discuss in the following how our models can capture these variations and what the reasons behind the renormalizations are.

From a qualitative point of view, Fig.~\ref{fig:spc}(b) answers the question of how the transition of a doubly-peaked conductivity in small crystals towards a single-peaked spectrum in large crystals appears. We would expect the peaks to merge or the oscillator strength of one of them to vanish as the crystal radius increases. We observe the second possibility, implying significant mixing of the states due to the Coulomb interaction. Note that extension of the basis towards the exact model does not significantly influence the observed ratio between the peaks. To address the renormalization of the oscillator strengths, let us consider relevant states in the strong confinement model. The ground $s$-symmetric state is constructed as a linear combination:
\begin{equation}
    \ket{\alpha_0} = a\ket{1s,1s}+b\ket{1p_x,1p_x}+b\ket{1p_y,1p_y}+b\ket{1p_z,1p_z}\,,
\end{equation}
where $a$, $b$ are the relevant coefficients $c_i$ from \eqref{eq:eigenstates}.
Upon excitation by an $x$-polarized field, the system undergoes a transition to an 
$x$-symmetric state within the linear response theory. Suitable states are:
\begin{eqnarray}
    \ket{\beta_+} &=& \phantom{-}c\ket{1s,1p_x}+ d\ket{1p_x,1s}\,,\\
    \ket{\beta_-} &=& -d\ket{1s,1p_x} + c\ket{1p_x,1s}\,,
\end{eqnarray}
where $c$, $d$ are once again the relevant coefficients $c_i$ from \eqref{eq:eigenstates}. The 
energy levels of the excited states are separated by:
\begin{equation}
    \Delta E = \sqrt{\Delta E_{k}^2+4|E_C|^2}\,,
    \label{eq:energy_sep}
\end{equation}
where $\Delta E_k$ is the splitting of the bare states (difference of their kinetic energies) and
$E_C$ is the off-diagonal interaction energy element in the relevant block of the Hamiltonian matrix
\begin{equation}
    E_C = \frac{-e^2}{4\pi\varepsilon} \left\langle 1p_x,1s \left| \frac{1}{|\bm{r}_e-\bm{r}_h|}
    \right| 1s,1p_x \right\rangle\,. \label{eq:corr_energy}
\end{equation}
The dipole moments (squared) of the two two-particle states' transitions can be expressed in terms 
of the dipole moments of the transitions between the single-particle states:
\begin{eqnarray}
    |\langle \beta_+ | d_x | \alpha_0 \rangle|^2 &=& e^2|(a-b)(c-d)\langle 1p_x|x|1s\rangle|^2\label{eq:dipRen1}\,,\\
    |\langle \beta_- | d_x | \alpha_0 \rangle|^2 &=& e^2|(a-b)(c+d)\langle 1p_x|x|1s\rangle|^2
    \label{eq:dipRen2}\,.
\end{eqnarray}
In case of no interaction $E_C=0$, the coefficients are $a=1,\; b=0,\; c=1,\; d=0$, and there is no difference in the dipole moments for the individual transitions of the electron or hole (provided the same shape of their respective wavefunctions). On the other hand, in the opposite case of Coulomb mixing of the states with $\Delta E_k=0$ (i.e. $m_e=m_h$), the relevant coefficients become $c=d$ and hence the lower-energy transition to the state $|\beta_+\rangle$ represented by the symmetric combination has zero dipole moment, while the dipole moment of the higher-energy transition to the state $|\beta_-\rangle$ with the anti-symmetric combination is augmented, fully in accord with the numerical data presented in Fig.~\ref{fig:spc}. Considering the scaling rule $E_C\propto A^{-1}$ and $\Delta E_k\propto A^{-2}$, the ratio $\Delta E_k/E_C\propto A^{-1}$ proceeds towards zero when increasing the crystal size, thus weakening the dipole moment of the transition to the state $\ket{\beta_+}$.

The nanocrystal size dependence of the ratio between the respective heights of hole- and electron-like peaks at resonances in the conductivity spectra significantly depends on the ratio between the electron and hole effective mass as depicted in Figs.~\ref{fig:spc}(e) and \ref{fig:lines}(d). In the most extreme case $m_e=m_h$ [relevant, for example, for PbS, Fig.~\ref{fig:lines}(a)], the lower-frequency peak disappears for any crystal size as noted above and shown in the plots. That means that the Coulomb interaction cannot, in general, be neglected even in the SCR limit  defined as $A \ll a_X$. To give a better estimate when the Coulomb interaction can be neglected, we take a numerically calculated value of the Coulomb coupling defined above $E_C={-}0.028e^2/\varepsilon A$ and consider for the kinetic energy:
\begin{equation}
    \Delta E_k=\frac{\hbar^2}{2}\left(k_{1p}^2-k_{1s}^2\right)\left|\frac{1}{m_e}-\frac{1}{m_h}\right|\,.
    \label{eq:Ek}
\end{equation}
For a spherical nanocrystal, $k_{1p}^2-k_{1s}^2\approx10/A^2$. Considering the Coulomb-coupled
states $\ket{1s,1p_x}$ and $\ket{1p_x,1s}$, it can be shown that the squared dipole moment renormalization
from Eqs.~(\ref{eq:dipRen1})--(\ref{eq:dipRen2}) is 10\% if $\Delta E_k\approx19|E_C|$. Substituting
for $E_C$ and Eq.~(\ref{eq:Ek}), we finally get the upper limit for the crystal radius to get
at most 10\% renormalization of the peak values:
\begin{equation}
    A_{10\%}=\frac{3\pi\hbar^2\varepsilon}{e^2}\left|\frac{1}{m_e}-\frac{1}{m_h}\right|\,.
    \label{eq:tenPerc}
\end{equation}
The above expression is similar to that for the exciton Bohr radius \eqref{eq:bohr_radius}, except for the numerical 
prefactor and a minus sign in the term with effective masses. It is then directly seen that Coulomb 
renormalizations become important even for $A\ll a_X$ if $m_e\approx m_h$ and hence equation 
(\ref{eq:tenPerc}) sets a more relevant criterion for neglection of Coulomb renormalizations for THz 
spectroscopy of nanocrystals. Note that $A_{10\%}<a_X$ is always satisfied. For the particular case of 
Fig.~\ref{fig:lines}(d) and the curves for the electron masses $m_e/M=0.25,0.35,0.45$, we get numerical 
values $A_{10\%}=7\ \mathrm{nm}, 3\ \mathrm{nm},1\ \mathrm{nm}$ and $a_X=18\ \mathrm{nm},15\ 
\mathrm{nm},13\ \mathrm{nm}$, respectively. These results mean that even for nanocrystals as small as roughly 10~nm 
radius, the hole resonance is significantly suppressed by the Coulomb interaction, while the electron 
resonance becomes amplified in all cases. In PbS, the values are $A_{10\%}=0$~nm and $a_X=13$~nm, and in GaAs 
$A_{10\%}=9$~nm, $a_X=12$~nm. This may cause difficulties in interpretation of the experimental data 
if the Coulomb interaction is not treated properly,
and therefore one should take the effect into consideration.

The strength of renormalization of resonances naturally depends on the number of states which are
taken into account in the calculation. This effect is most prominent in large crystals since the Coulomb
interaction dominates over the kinetic energy, as seen in Fig.~\ref{fig:lines}(d) where the ratios of
electron- and hole-like resonance magnitudes multiplied by the ratio $m_e/m_h$ are plotted by dotted lines,
a result of SCR calculation with 729 basis states. In small crystals, the relative correction is
not too large and the above estimate of the $A_{10\%}$ criterion remains valid. Results of calculations
also show that the peak renormalization is well estimated by the SCR model with the minimal basis unless
the crystal radius exceeds the exciton Bohr radius.
In indirect gap semiconductors 
[Fig.~\ref{fig:lines}(b)], similar considerations can be done, however, the situation is complicated 
by the degeneracy due to the multiple valleys in the conduction band and their effective mass anisotropy.
As a result, significant modifications of the spectra are already 
observable for crystal radius 10~nm as seen in Fig.~\ref{fig:lines}(b).

\subsection{Indirect-gap semiconductors}

Indirect-gap semiconductors can provide highly anisotropic conduction band minima outside the $\Gamma$ point of the Brillouin zone such that the electron effective mass in one direction is lower than that of a hole, as usual, but in another direction the electron can be considerably heavier than valence holes. We take silicon nanocrystals of radius 10~nm as an example, whose conductivity spectrum is already depicted in Fig.~\ref{fig:lines}(b). The highest-energy (heavy) holes have an effective mass $0.49m_0$. The conduction band minima lie in six X valleys with the longitudinal effective mass $0.98m_0$ and the transverse effective mass $0.19m_0$. The lowest-frequency bare hole transition between the $1s$ and $1p$ states (transition between the ground and the first excited bare electron-hole state with excited hole) appears at 1.94~THz and the bare electron transition frequencies (longitudinal and transverse) are 0.97~THz and 5.01~THz, respectively. Coulomb interaction couples $p$-like states $\big|1s^{X_j}1p^\Gamma_{x,y,z}\big\rangle\leftrightarrow\big|1p^{X_j}_{x,y,z} 1s^\Gamma\big\rangle$ within each of the $X_j$ valleys (intra-valley coupling), conserving the $p$-state polarization. The hole states are localized within the $\Gamma$ point. Intra-valley interaction thus shifts energies and dipole moments of the dipole transition in each of the valleys separately and therefore we expect bleaching of the hole resonance paired with the transverse electron while the electron-like response is diminished in the case of the longitudinal electron. Energy renormalization lifts the degeneracy of the hole-like resonance, thus leading to four pronounced peaks in the conductivity spectra.

Quantum confinement causes relaxation of the momentum conservation
law, and the Coulomb interaction therefore results in inter-valley coupling whose relative importance,
as compared to the intra-valley scattering, increases when reducing the crystal size. The Coulomb interaction
therefore couples electron-hole pairs within an extended system of states in small crystals
beyond simple intra-valley coupling, and we expect that the doubly degenerate
transverse electron-like state splits further. This is seen in Fig.~\ref{fig:lines}(b) as the splitting of
the highest-energy resonance at $\sim$6~THz while the longitudinal electron-like resonance at $\sim$0.97~THz
vanishes completely.

Lifting of the hole resonance degeneracy, and splitting of the transverse-electron state due to inter-valley Coulomb
scattering can be identified in Fig.~\ref{fig:lines}(c) where the overall THz response of 10~nm Si nanocrystals
is replotted from Fig.~\ref{fig:lines}(b) (black line) and the spectrum for 12.5~nm nanocrystals is added for comparison (green line). Additional lines depict the response spectrum if only electrons (blue line) or holes (red line) were coupled to the probing electromagnetic field. The dominant component (electron or hole) then reveals whether a particular resonance in the overall spectrum has its origin in the electron or hole state. The bare longitudinal
electron resonance at 0.97~THz renormalizes to 1.2~THz while the two resonances at 2.4 and 3.5 THz are hole-like,
indicating splitting and shift of the bare 1.94~THz resonance. Interestingly, the splitting does not
depend strongly on the crystal size; see the green line for 12.5~nm crystal. A crossover from non-interacting to
fully interacting system, in which case the splitting of the electron-like and hole-like resonances
can be well identified, is seen in Supplemental Material \cite{supplemental}, Sec. I and Fig.~S1 where we plot the normalized conductivity
while changing the Coulomb interaction strength via a modification of the material relative permittivity.

\subsection{Weak confinement}

Renormalization of the dipole moments can also be viewed from another perspective. Our WCR
model describes only the resonances which come from the exciton internal degrees of freedom since they
are decoupled from the COM motion by factorization of the wavefunction (see the Appendix~\ref{app:a}).
In a kinetic picture, the COM motion is coupled to the internal degrees of freedom each time
one of the particles hits a crystal wall. As a consequence, the COM
couples to the external field, even though the exciton is electrically neutral, causing the appearance of
an additional, lower-frequency resonance in the response function. When collisions become
rare in a large crystal, the lower-frequency resonance bleaches as the coupling vanishes.
From this point of view, the condition for the WCR $A\gg a_X$ is in agreement with the requirement 
on the weak exciton scattering on the crystal walls: the probability of finding an exciton closer to 
the crystal wall than a half of the exciton Bohr radius is less than 10\% for $A \gtrsim 10a_X$.

Although the WCR and SCR models do not cover the region of crystal radii $a_X < A < 10a_X$ (ICR),
our WCR model can be extended beyond the upper limit for the crystal radius at least as an estimate 
of the major resonance in the spectrum. We argue that the exciton energy is well defined even in 
small crystals where the dead layer
model fails due to a fixed radius of the exciton. The situation is best displayed in Fig.~\ref{fig:mmt}(a)
by comparing the excitation energy of the electron-hole pair plotted by red and orange lines.
While the red line (WCR model) approaches
the SCR model in small crystals, the dead layer model diverges. On the other side of the
crystal size, the WCR model correctly reaches the exciton bulk value. The correct
asymptotic behavior of the WCR model is further illustrated in Figs.~\ref{fig:mmt}(b)-(c)
where both the dipole moment of the major resonance and the oscillator strength smoothly connect the
SCR on one side with the bulk exciton limit on the other. 
The SCR model can be hypothetically extended over the whole ICR region, assuming sufficiently
large set of basis electron-hole wavefunctions. The estimate of the necessary number of states
is derived in Appendix~\ref{app:b}, resulting in the basis size of approx. $2\times10^4$ states (GaAs, $A=100\ \mathrm{nm} < 10 a_X$)
which is too large for a numerical evaluation.

Interestingly, the dipole moment in Fig.~\ref{fig:mmt}(b) has a maximum near the crystal radius equal
to the Bohr radius of the $n=2$ exciton state. At this confinement size, the $(2p)_w$ 
wavefunctions are strongly shrunken by the crystal walls, increasing the spatial overlap with 
the $(1s)_w$ wavefunction, which remains relatively untouched.

It is worth noting that a proper choice of the function $\varrho(r)$ in the WCR model 
is crucial for the way in which the predictions of the WCR and SCR models meet at small crystal radii. 
We observe in Fig.~\ref{fig:mmt}(b) that the dipole moments join smoothly as the crystal size 
decreases all the way down to the region of validity of the SCR model. This is possible only when 
we use the definition (\ref{eq:radius_rho}) otherwise the red and the blue lines either intersect or 
connect sharply in the limit of zero crystal radius.

\subsection{Thermal fluctuations}

The state of the exciton to be probed in an experiment will largely depend on the particular 
experimental conditions. In order to predict whether the spectral features described in the SCR
and WCR models above will be present in the measured THz response at given temperature, 
the thermal energy $k_B T$ should be compared to the relevant energy scales of the respective 
models. In the case of the SCR model, the relevant quantity is $E_C$, see Eq. 
\eqref{eq:corr_energy}, which is responsible for the correlations between the phase of the electron 
and hole states. In the case of the WCR model, we consider binding energies $u_0, 
u_1$:
\begin{eqnarray}
    u_0 = E_{(1s)_w} - E_0^{(H)}\,,&\\
    u_1 = E_{(2p)_w} - E_0^{(H)}\,,&
    \label{eq:bind_energy}
\end{eqnarray}
which separate the bound states (the ground state $(1s)_w$ and the first excited state $(2p)_w$, 
respectively) from the quasi-continuum of uncorrelated (ionized) states of the electron-hole pair 
(Hartree ground state energy $E_0^{(H)}$). Fig.~\ref{fig:mmt}(d) offers a comparison of the 
Coulomb correlation energy $E_C$ and the binding energy $u_0$ with the thermal energy $k_B T$.
The plot further illustrates under which experimental conditions (temperature and crystal size) the
Coulomb correlations would be observable.

For sufficiently low temperatures, i.e. $k_{B}T < |E_C|$ or $k_{B}T < |u_0|,|u_1|$, we expect high 
coherence of the system, and hence the correlations between the electron and the hole induced by the 
Coulomb interaction should be preserved. These correlations include the bound excitonic states in 
the WCR, resulting in the discrete response spectrum, and the Coulomb-induced energy shifts
and renormalization of dipole moments in the electron and hole excitation spectrum in the SCR due to 
the correlated phase of the electron and hole excited states.

The situation will be different in the case of elevated temperatures. Within the 
WCR, the thermal fluctuations of $k_{B}T > |u_0|,|u_1|$ lead to ionization of the exciton. The 
electron and hole will then occupy the quasi-continuum of states of the weakly confined individual 
electron and hole. A Drude-like conductivity contribution is expected to define the THz response 
spectrum.

In the SCR, there is no continuum of ionized states. The large thermal energy 
$k_{B}T > E_C$ causes decoherence and loss of correlation between the electron and the hole. In order 
to capture the suppressed correlation in the model, it is justified to neglect the off-diagonal 
elements of the Hamiltonian \eqref{eq:ham_gen} expressed in the basis 
of bare two-particle states, resulting in a mean-field single-particle approximation. 
From the spectroscopic point of view, however, the most prominent effect at high temperature would be a broadening of the spectral lines, controlled by the momentum relaxation rate $\gamma$ in Eq. \eqref{eq:conductivity}, making an observation of the decay of correlations and
thus eventual spectral shifts and peak renormalizations challenging.
Furthermore, provided sufficient thermal energy $k_{B}T > \Delta E_{k}$, there is a nonzero 
probability of finding the electron and hole in higher excited states to be probed by the THz field 
Accordingly, additional fundamental transitions contribute to the observed response spectrum
while the low-frequency resonances diminish due to similar populations of initial and final states which are close in energy. 

We expect that the amount of increase of the parameter $\gamma$ with temperature is sample-dependent as it is 
strongly affected by the nanocrystal geometry and surface quality. The surface is of a major importance 
in small nanocrystals owing to which we expect an enhanced value of $\gamma$ with respect to its bulk 
value and also its weak temperature dependency. The value of $\gamma$ should converge to the bulk one 
with increasing the nanocrystal size, making its temperature dependency more prominent.

We observe in Fig.~\ref{fig:mmt}(d) that the relevant quantities $u_0$ and $E_C$ meet in the region 
of crystal sizes corresponding to ICR. The two lines intersect at approximately $A=2.25a_X$ 
($a_X=12$ nm in GaAs, $a_X=4$ nm in GaN.) As stated above, in ICR, both the WCR and the SCR models 
face serious limitations. The WCR model neglects the coupling of COM to the external field, and the 
SCR model places enormous demands on the size of the computational basis. Nevertheless, if we wish 
to interpolate the response in ICR using the models, $A > 2.25a_X$ gives a terminal crystal radius, 
above which the WCR model could be successfully applied to describe at least the major resonance. 
Above this limit, the bound, yet shrunken, Wannier-like exciton state is stable. Below this limit, 
the shrinkage leads to a large kinetic energy, rendering the bound state unstable, in which case one 
should resort to the SCR model instead.


\section{\label{sec:conclusion}Conclusion}

We have addressed the problem of an electron-hole pair confined in a nanocrystal from the point of view
of THz spectroscopy. Our analysis uncovers significant variations in conductivity spectra as a result of 
the interplay between the electron-hole Coulomb interaction and particle confinement. We show that
substantial renormalizations appear already if the separation of the confined two-particle states is
comparable to the Coulomb interaction energy and we propose relevant quantum-mechanical models
beyond the mean field theory in order to capture this effect.
We derive two models, which are applicable to different nanocrystal size scales: while the strong
confinement model is applicable to crystals with radius smaller than or comparable to the exciton Bohr radius, the weak
confinement model is valid for crystals much larger than the exciton Bohr radius. Using the strong confinement
model, we demonstrate the transition of doubly-peaked ac conductivity in small crystals to single-peaked conductivity
in larger crystals. This behavior is interpreted in terms of mixing of two-particle states by Coulomb
electron-hole interaction. We determine the limiting crystal radius $A_{10\%}$
below which the electron-hole pair may be regarded as independent particles. Interestingly, this quantity
can be considerably smaller than the exciton Bohr radius $a_X$ and reaches zero in materials with equal
electron and hole mass (for example, PbS). We show that the newly proposed weak confinement model smoothly connects
the bulk and the strongly confined limits of the system in terms of resonant frequency and dipole moment of the major inter-subband transition as well as oscillator strength, and thus can be used down to a crystal size comparable to the exciton Bohr radius. The weak confinement model balances the kinetic energies of the relative and center-of-mass modes for each state, thus making observation of energy renormalizations of highly excited excitonic states possible. It is therefore directly applicable to describe Rydberg states of confined excitons.


\begin{acknowledgments}
  This work was supported by the Czech Science Foundation (Project No. 23-05640S) and by 
  TERAFIT Project No. CZ.02.01.01/00/22\_008/0004594 funded by OP JAK, call Excellent Research.
  We also acknowledge the financial support by the Charles University grant SVV–2025–260836.
  
\end{acknowledgments}

\appendix
\section{\label{app:derivation}Derivation of weak confinement model extension\label{app:a}}

The Schrödinger equation with the weak confinement Hamiltonian \eqref{eq:ham_weak_conf} reads:
\begin{multline}
    \left[ -\frac{\hbar^2}{2M}\nabla_R^2 +V\big(R+\varrho(r)\big) 
  -\frac{\hbar^2}{2\mu}\nabla_r^2  -\frac{e^2}{4\pi \varepsilon r} \right] \times \\
    \times \Phi(\bm{R},\bm{r})
    = E\; \Phi(\bm{R},\bm{r})\,.
\end{multline}
The equation above is not separable in terms of coordinates $\vec{R}$ and $\vec{r}$ due to the term
$V(R+\varrho(r))$. Nevertheless, we assume a separated form of the solution:
\begin{equation}
    \Phi(\bm{R},\bm{r}) = \Psi(\bm{R};r)\psi(\bm{r})
\end{equation}
in order to find at least an approximate solution. This particular approximation is justified in the Supplemental Material \cite{supplemental}, Section II. This step will break the correlations between
the center-of-mass motion and the internal degrees of freedom with the consequences outlined in the discussion
section. The term $V(R+\varrho(r))$ will be kept in the equations, allowing us to balance the kinetic and
potential energy between the separated degrees of freedom. The separated form of the equation is
\begin{eqnarray}
    \left[ -\frac{\hbar^2}{2M}\nabla_R^2 +V\big( R+\varrho(r) \big) \right] \Psi(\bm{R};r) &=& \chi(r) \Psi(\bm{R};r)\label{ap3}\\
    \left[ -\frac{\hbar^2}{2\mu}\nabla_r^2 -\frac{1}{4\pi\varepsilon}\frac{e^2}{r} -E \right] \psi(\bm{r}) &=& -\chi(r) \psi(\bm{r})\,.\quad\quad\quad
    \label{ap4}
\end{eqnarray}
The center-of-mass equation is solved first, in order to find $\chi(r)$. The Hamiltonian of the 
center-of-mass is the one of the particle-in-a-box with well-known solutions for the wavefunction 
$\Psi(\bm{R};r)$ and kinetic energy $\chi(r)$, with $r$ being treated as a parameter:
\begin{eqnarray}
    \Psi_{NLM}(\bm{R};r) &=& \mathcal{N}_{NL} j_L\left(k_{NL}(r)R\right) Y_L^M(\Theta,\Phi)\\
    \chi_{NL}(r) &=& \frac{\hbar^2 k_{NL}^2(r)}{2M}\,,
\end{eqnarray}
where $j_L$ are the spherical Bessel functions, $Y_L^M$ the spherical harmonics, $\mathcal{N}_{NL}$ 
the normalization constants and the wavevectors $k_{NL}$ satisfy the boundary condition
\begin{equation}
    j_L\big( k_{NL}(r) \left[ A-\varrho(r) \right] \big) = 0\,.
\end{equation}
The ground-state ($N=1$, $L=0$) energy of the center-of-mass motion is
\begin{equation}
    \chi_{1,0}(r) = \frac{\hbar^2}{2M}\left(\frac{\pi}{A-\varrho(r)}\right)^2\,.
    \label{eq:com_ground_state_app}
\end{equation}
The remaining equation for the relative part has the form of a hydrogen atom with an additional potential 
energy term $\chi_{NL}(r)$, which has a $r^{-2}$ singularity at 
$r=\min\left(\frac{m_e}{\mu}A;\frac{m_h}{\mu}A\right)$, due to the confinement 
potential:
\begin{equation}
    \left[ -\frac{\hbar^2}{2\mu}\nabla_r^2 -\frac{1}{4\pi\varepsilon}\frac{e^2}{r} +\chi_{NL}(r) 
  \right] \psi(\bm{r}) = E \psi(\bm{r})\,.
\end{equation}
The solution is to be found in the form
\begin{equation}
    \psi_{nlm}(\bm{r}) = \mathcal{N}_{nl} \mathcal{R}_{nl}(r) Y_l^m(\theta,\phi)\,,
\end{equation}
where $\mathcal{R}_{nl}$ are the solutions of the radial equation
\begin{multline}
    \left[ -\frac{\hbar^2}{2\mu}\frac{\partial^2}{\partial r^2} 
    -\frac{\hbar^2}{2\mu}\frac{2}{r}\frac{\partial}{\partial r} + 
    \frac{\hbar^2}{2\mu}\frac{l(l+1)}{r^2}-\right.\\
    \left.- \frac{1}{4\pi \varepsilon}\frac{e^2}{r} + \chi_{NL}(r) \right]
    \mathcal{R}(r) = E\; \mathcal{R}(r)\,,
    \label{eq:radial_eq_app}
\end{multline}
which also yields the total energy $E$ of the two-particle system.

\section{Number of states in strong confinement model\label{app:b}}

The exciton wave function can be regarded as a linear combination of bare-electron hole pairs where
the mixing of the bare states has the origin in the Coulomb interaction. The characteristic
interaction energy will be estimated using Eq.~(\ref{eq:corr_energy}) as $E_C=-0.028e^2/\varepsilon A$.
Only those states which have a bare energy less than or comparable to $|E_C|$ will contribute to the 
lowest-energy exciton states, hence we may restrict ourselves to $\hbar^2k_{nl}^2/2m_j\leq 10|E_C|$
[cf. Eq.~(\ref{eq:bareEnergy})] for each
of the electron/hole (considering $n,l\gg1$, thus neglecting the ground state energy). The values
$k_{nl}$ are defined so that $j_l(k_{nl}A)=0$, therefore it holds that $k_{nl}=x_{nl}/A$ where $x_{nl}$ is the
$n$-th root of the spherical Bessel function $j_l(x)$, to which we apply the asymptotic formula $x_{nl}\approx\pi(n+l/2)$.
The single-particle states which contribute to the exciton wavefunction are then characterized by
the inequality:
\begin{equation}
    n+\frac{l}{2}\leq\frac{e}{\hbar\pi}\sqrt{\frac{0.56m_jA}{\varepsilon}}=N_j\,,
\end{equation}
where we defined $N_j$. The upper bound for the value of $l$ is $2N_j$ and for a fixed $l$,
the number of different values of $n$ is $N_j-l/2$. Taking into account the $2l+1$ degeneracy for each $l$,
the total number of single-particle states that must be taken into account is:
\begin{equation}
    \sum_{l=0}^{2N_j}(2l+1)\Big(N_j-\frac{l}{2}\Big)\approx\frac{4}{3}N_j^3
\end{equation}
for large $N_j$. The total number of necessary two-particle states is then $N_{\mathrm{TOT}}\approx 2(N_eN_h)^3$ which
yields $N_{\mathrm{TOT}}\approx500$ in GaAs for $A=30$~nm and $N_{\mathrm{TOT}}\approx2\times10^4$
for $A=100$~nm.


\bibliography{main_arxiv}

@article{anantharaman2021,
  Author = {Anantharaman, Surendra B. and Jo, Kiyoung and Jariwala, Deep},
  Title = {Exciton-Photonics: From Fundamental Science to Applications},
  Journal = {ACS Nano},
  Year = {2021},
  Volume = {15},
  Number = {8},
  Pages = {12628-12654},
  Month = {AUG 24},
  DOI = {10.1021/acsnano.1c02204},
  EarlyAccessDate = {JUL 2021},
  ISSN = {1936-0851},
  EISSN = {1936-086X}
}

@article{efros2021,
  Author = {Efros, Alexander L. and Brus, Louis E.},
  Title = {Nanocrystal Quantum Dots: From Discovery to Modern Development},
  Journal = {ACS Nano},
  Year = {2021},
  Volume = {15},
  Number = {4},
  Pages = {6192-6210},
  Month = {APR 27},
  DOI = {10.1021/acsnano.1c01399},
  EarlyAccessDate = {APR 2021},
  ISSN = {1936-0851},
  EISSN = {1936-086X}
}

@article{harankahage2021,
  Author = {Harankahage, Dulanjan and Cassidy, James and Yang, Mingrui and
    Porotnikov, Dmitry and Williams, Maia and Kholmicheva, Natalia and
    Zamkov, Mikhail},
  Title = {Quantum Computing with Exciton Qubits in Colloidal Semiconductor
    Nanocrystals},
  Journal = {J. Phys. Chem. C},
  Year = {2021},
  Volume = {125},
  Number = {40},
  Pages = {22195-22203},
  Month = {OCT 14},
  DOI = {10.1021/acs.jpcc.1c05009},
  EarlyAccessDate = {OCT 2021},
  ISSN = {1932-7447},
  EISSN = {1932-7455}
}

@article{kaindl2003,
  Author = {Kaindl, R A and Carnahan, M A and Hägele, D and Lövenich, R and Chemla, D S},
  Title = {Ultrafast terahertz probes of transient conducting and insulating phases
    in an electron-hole gas},
  Journal = {Nature},
  Year = {2003},
  Volume = {423},
  Number = {6941},
  Pages = {734-738},
  Month = {JUN 12},
  DOI = {10.1038/nature01676},
  ISSN = {0028-0836},
  EISSN = {1476-4687}
}

@article{kuzel2019,
  Author = {Ku\v{z}el, Petr and N\v{e}mec, Hynek},
  Title = {Terahertz Spectroscopy of Nanomaterials: a Close Look at Charge-Carrier
    Transport},
  Journal = {Adv. Opt. Mater.},
  Year = {2020},
  Volume = {8},
  Number = {3, SI},
  Month = {FEB},
  DOI = {10.1002/adom.201900623},
  Article-Number = {1900623},
  ISSN = {2195-1071}
}

@article{leinss2008,
  Author = {Leinss, S. and Kampfrath, T. and von Volkmann, K. and Wolf, M. and
    Steiner, J. T. and Kira, M. and Koch, S. W. and Leitenstorfer, A. and
    Huber, R.},
  Title = {Terahertz Coherent Control of Optically Dark Paraexcitons in
    {Cu\textsubscript{2}O}},
  Journal = {Phys. Rev. Lett.},
  Year = {2008},
  Volume = {101},
  Number = {24},
  Month = {DEC 12},
  DOI = {10.1103/PhysRevLett.101.246401},
  Article-Number = {246401},
  ISSN = {0031-9007}
}

@article{li2020,
  Author = {Li, Xinwei and Yoshioka, Katsumasa and Zhang, Qi and Peraca, Nicolas
    Marquez and Katsutani, Fumiya and Gao, Weilu and Noe, II, G. Timothy and
    Watson, John D. and Manfra, Michael J. and Katayama, Ikufumi and Takeda,
    Jun and Kono, Junichiro},
  Title = {Observation of Photoinduced Terahertz Gain in {GaAs} Quantum Wells:
    Evidence for Radiative Two-Exciton-to-Biexciton Scattering},
  Journal = {Phys. Rev. Lett.},
  Year = {2020},
  Volume = {125},
  Number = {16},
  Month = {OCT 16},
  DOI = {10.1103/PhysRevLett.125.167401},
  Article-Number = {167401},
  ISSN = {0031-9007},
  EISSN = {1079-7114}
}

@article{pushkarev2022,
  author = {Pushkarev, Vladimir and N\v{e}mec, Hynek and Paingad, Vaisakh C. and Ma\v{n}\'{a}k, Jan and Jurka,
    Vlastimil and Nov\'{a}k, V\'{\i}t and Ostatnick\'{y}, Tom\'{a}\v{s} and Ku\v{z}el, Petr},
  title = {Charge Transport in Single-Crystalline {GaAs} Nanobars: Impact of Band Bending Revealed by Terahertz Spectroscopy},
  journal = {Advanced Functional Materials},
  volume = {32},
  number = {5},
  pages = {2107403},
  doi = {https://doi.org/10.1002/adfm.202107403},
  year = {2022}
}

@article{sanderson2019,
  Author = {Sanderson, William M. and Wang, Fudong and Buhro, William E. and Loomis,
    Richard A.},
  Title = {Long-Lived 1D Excitons in Bright {CdTe} Quantum Wires},
  Journal = {J. Phys. Chem. C},
  Year = {2019},
  Volume = {123},
  Number = {5},
  Pages = {3144-3151},
  Month = {FEB 7},
  DOI = {10.1021/acs.jpcc.8b09588},
  ISSN = {1932-7447},
  EISSN = {1932-7455}
}

@article{nemec2009,
  title = {Far-infrared response of free charge carriers localized in semiconductor nanoparticles},
  author = {N\v{e}mec, Hynek and Ku\v{z}el, Petr and Sundstr\"om, Villy},
  journal = {Phys. Rev. B},
  volume = {79},
  issue = {11},
  pages = {115309},
  numpages = {7},
  year = {2009},
  month = {Mar},
  publisher = {American Physical Society},
  doi = {10.1103/PhysRevB.79.115309}
}

@article{quick2022,
  Author = {Quick, Michael T. and Ayari, Sabrine and Owschimikow, Nina and Jaziri,
    Sihem and Achtstein, Alexander W.},
  Journal = {ACS Appl. Nano Mater.},
    Year = {2022},
  Volume = {5},
  Number = {6},
  Pages = {8306-8313},
  Month = {JUN 24},
  DOI = {10.1021/acsanm.2c01385},
  title = {Quantum Nature of {THz} Conductivity: Excitons, Charges, and Trions in {2D} Semiconductor Nanoplatelets and Implications for {THz} Imaging and Solar Hydrogen Generation}
}

@article{pushkarev2017,
  title = {Quantum behavior of terahertz photoconductivity in silicon nanocrystals networks},
  author = {Pushkarev, V. and Ostatnick\'y, T. and N\v{e}mec, H. and Chlouba, T. and Troj\'anek,
    F. and Mal\'y, P. and Zacharias, M. and Gutsch, S. and Hiller, D. and Ku\v{z}el, P.},
  journal = {Phys. Rev. B},
  volume = {95},
  issue = {12},
  pages = {125424},
  numpages = {9},
  year = {2017},
  month = {Mar},
  publisher = {American Physical Society},
  doi = {10.1103/PhysRevB.95.125424}
}

@article{steinleitner2017,
  Author = {Steinleitner, Philipp and Merkl, Philipp and Nagler, Philipp and
    Mornhinweg, Joshua and Schueller, Christian and Korn, Tobias and
    Chernikov, Alexey and Huber, Rupert},
  Title = {Direct Observation of Ultrafast Exciton Formation in a Monolayer of
    {WSe\textsubscript{2}}},
  Journal = {Nano Lett.},
  Year = {2017},
  Volume = {17},
  Number = {3},
  Pages = {1455-1460},
  Month = {MAR},
  DOI = {10.1021/acs.nanolett.6b04422},
  ISSN = {1530-6984},
  EISSN = {1530-6992}
}

@article{tagarelli2023,
  Author = {Tagarelli, Fedele and Lopriore, Edoardo and Erkensten, Daniel and
    Perea-Causin, Rauel and Brem, Samuel and Hagel, Joakim and Sun, Zhe and
    Pasquale, Gabriele and Watanabe, Kenji and Taniguchi, Takashi and Malic,
    Ermin and Kis, Andras},
  Title = {Electrical control of hybrid exciton transport in a {van der Waals}
    heterostructure},
  Journal = {Nat. Photonics},
  Year = {2023},
  Volume = {17},
  Number = {7},
  Pages = {615+},
  Month = {JUL},
  DOI = {10.1038/s41566-023-01198-w},
  EarlyAccessDate = {APR 2023},
  ISSN = {1749-4885},
  EISSN = {1749-4893}
}

@article{efros1982,
  Author = {Efros, Alexei L and Efros, {\relax Al}exander L},
  Title = {Interband absorption of light in a semiconductor sphere},
  Journal = {Sov. Phys. Semicond.},
  Year = {1982},
  Volume = {16},
  Number = {7},
  Pages = {772-775},
  ISSN = {0038-5700}
}

@book{banyai1993,
  title={Semiconductor Quantum Dots},
  author={Banyai, L. A. and Koch, S. W.},
  isbn={9789814504232},
  series={World Scientific Series on Atomic, Molecular and Optical Physics},
  url={https://books.google.cz/books?id=BrzsCgAAQBAJ},
  year={1993},
  publisher={World Scientific Publishing Company},
  pages={37--62}
}

@article{rossi2020,
  Author = {Rossi, Daniel and Liu, Xiaohan and Lee, Yangjin and Khurana, Mohit and
    Puthenpurayil, Joseph and Kim, Kwanpyo and Akimov, V, Alexey and Cheon,
    Jinwoo and Son, Dong Hee},
  Title = {Intense Dark Exciton Emission from Strongly Quantum-Confined
    {CsPbBr\textsubscript{3}} Nanocrystals},
  Journal = {Nano Lett.},
  Year = {2020},
  Volume = {20},
  Number = {10},
  Pages = {7321-7326},
  Month = {OCT 14},
  DOI = {10.1021/acs.nanolett.0c02714},
  ISSN = {1530-6984},
  EISSN = {1530-6992}
}

@article{haldar2017,
  Author = {Haldar, S. and Dixit, V. K. and Vashisht, Geetanjali and Khamari,
    Shailesh Kumar and Porwal, S. and Sharma, T. K. and Oak, S. M.},
  Title = {Effect of carrier confinement on effective mass of excitons and
    estimation of ultralow disorder in
    {Al\textsubscript{x}Ga\textsubscript{1-x}As/GaAs} quantum wells by
    magneto-photoluminescence},
  Journal = {Sci. Rep.},
  Year = {2017},
  Volume = {7},
  Month = {JUL 7},
  DOI = {10.1038/s41598-017-05139-w},
  Article-Number = {4905},
  ISSN = {2045-2322}
}

@article{quick2023,
  Author = {Quick, Michael T. and Wach, Quentin and Owschimikow, Nina and Achtstein,
    Alexander W.},
  Title = {{THz} Response of Charge Carriers in Nanoparticles: Microscopic Master
    Equations Reveal an Unexplored Equilibration Current and Nonlinear
    Mobility Regimes},
  Journal = {Adv. Photon. Res.},
  Year = {2023},
  Volume = {4},
  Number = {2},
  Month = {2023 FEB},
  DOI = {10.1002/adpr.202200243},
  EarlyAccessDate = {DEC 2022},
  ISSN = {2699-9293}
}

@article{ostatnicky2019,
  Author = {Ostatnick\'{y}, T.},
  Title = {Linear {THz} conductivity of nanocrystals},
  Journal = {Opt. Express},
  Year = {2019},
  Volume = {27},
  Number = {5},
  Pages = {6083-6088},
  Month = {MAR 4},
  DOI = {10.1364/OE.27.006083},
  ISSN = {1094-4087}
}

@article{ostatnicky2018,
  Author = {Ostatnick\'{y}, T. and Pushkarev, V. and N\v{e}mec, H. and Ku\v{z}el, P.},
  Title = {Quantum theory of terahertz conductivity of semiconductor nanostructures},
  Journal = {Phys. Rev. B},
  Year = {2018},
  Volume = {97},
  Number = {8},
  Month = {FEB 20},
  DOI = {10.1103/PhysRevB.97.085426},
  Article-Number = {085426},
  ISSN = {2469-9950},
  EISSN = {2469-9969}
}

@article{ziemkiewicz2020,
  Author = {Ziemkiewicz, David and Karpi\'{n}ski, Karol and Czajkowski, Gerard and
    Zieli\'{n}ska-Raczy\'{n}ska, Sylwia},
  Title = {Excitons in {Cu\textsubscript{2}O}: From quantum dots to bulk crystals and
    additional boundary conditions for {Rydberg} exciton-polaritons},
  Journal = {Phys. Rev. B},
  Year = {2020},
  Volume = {101},
  Number = {20},
  Month = {MAY 4},
  DOI = {10.1103/PhysRevB.101.205202},
  Article-Number = {205202},
  ISSN = {2469-9950},
  EISSN = {2469-9969}
}

@article{bellessa1999,
  Author = {Bellessa, J and Combescot, M},
  Title = {Size dependance of impurity levels in quantum dots: exact versus
    variational results},
  Journal = {Solid State Com.},
  Year = {1999},
  Volume = {111},
  Number = {5},
  Pages = {275-280},
  DOI = {10.1016/S0038-1098(99)00147-7},
  ISSN = {0038-1098}
}

@article{chuu1992,
  Author = {Chuu, D S and Hsiao, C M and Mei, W N},
  Title = {Hydrogenic impurity states in quantum dots and quantum wires},
  Journal = {Phys. Rev. B},
  Year = {1992},
  Volume = {46},
  Number = {7},
  Pages = {3898-3905},
  Month = {AUG 15},
  DOI = {10.1103/PhysRevB.46.3898},
  ISSN = {0163-1829}
}

@article{dresselhaus1956,
  Author = {Dresselhaus, G},
  Title = {Effective mass approximation for excitons},
  Journal = {J. Phys. Chem. Solids},
  Year = {1956},
  Volume = {1},
  Number = {1-2},
  Pages = {14-22},
  DOI = {10.1016/0022-3697(56)90004-X},
  ISSN = {0022-3697}
}

@article{lage1991,
  Author = {Lage, H and Heitmann, D and Cingolani, R and Grambow, P and Ploog, K},
  Title = {Center-of-mass quantization of excitons in {GaAs} quantum-well wires},
  Journal = {Phys. Rev. B},
  Year = {1991},
  Volume = {44},
  Number = {12},
  Pages = {6550-6553},
  Month = {SEP 15},
  DOI = {10.1103/PhysRevB.44.6550},
  ISSN = {1098-0121}
}

@article{lee2018,
  Author = {Lee, Woojin and Kim, Minwoo and Yang, Hanyi and Kyhm, Kwangseuk and
    Murayama, Akihiro and Kheng, Kuntheak and Mariette, Henri and Dang, Le
    Si},
  Title = {Two-dimensional Nature of Center-of-mass Excitons Confined in a Single
    {CdMnTe/CdTe/CdMnTe} Heterostructure},
  Journal = {Curr. Opt. Photon.},
  Year = {2018},
  Volume = {2},
  Number = {6},
  Pages = {589-594},
  Month = {DEC},
  DOI = {10.3807/COPP.2018.2.6.589},
  ISSN = {2508-7266},
  EISSN = {2508-7274}
}

@article{chernikov2014,
  Author = {Chernikov, Alexey and Berkelbach, Timothy C. and Hill, Heather M. and
    Rigosi, Albert and Li, Yilei and Aslan, Ozgur Burak and Reichman, David
    R. and Hybertsen, Mark S. and Heinz, Tony F.},
  Title = {Exciton Binding Energy and Nonhydrogenic {Rydberg} Series in Monolayer
    {WS\textsubscript{2}}},
  Journal = {Phys. Rev. Lett.},
  Year = {2014},
  Volume = {113},
  Number = {7},
  Month = {AUG 13},
  DOI = {10.1103/PhysRevLett.113.076802},
  Article-Number = {076802},
  ISSN = {0031-9007},
  EISSN = {1079-7114}
}

@article{kazimierczuk2014,
  Author = {Kazimierczuk, T. and Froehlich, D. and Scheel, S. and Stolz, H. and
    Bayer, M.},
  Title = {Giant {Rydberg} excitons in the copper oxide {Cu\textsubscript{2}O}},
  Journal = {Nature},
  Year = {2014},
Volume = {514},
  Number = {7522},
  Pages = {343-347},
  Month = {OCT 16},
  DOI = {10.1038/nature13832},
  ISSN = {0028-0836},
  EISSN = {1476-4687}
}

@article{orfanakis2022,
  Author = {Orfanakis, Konstantinos and Rajendran, Sai Kiran and Walther, Valentin
    and Volz, Thomas and Pohl, Thomas and Ohadi, Hamid},
  Title = {{Rydberg} exciton-polaritons in a {Cu\textsubscript{2}O} microcavity},
  Journal = {Nat. Mater.},
  Year = {2022},
  Volume = {21},
  Number = {7},
  Pages = {767+},
  Month = {2022 JUL},
  DOI = {10.1038/s41563-022-01230-4},
  EarlyAccessDate = {APR 2022},
  ISSN = {1476-1122},
  EISSN = {1476-4660}
}

@article{orfanakis2021,
  Author = {Orfanakis, Konstantinos and Rajendran, Sai Kiran and Ohadi, Hamid and
    Zielinska-Raczynska, Sylwia and Czajkowski, Gerard and Karpinski, Karol
    and Ziemkiewicz, David},
  Title = {Quantum confined {Rydberg} excitons in {Cu\textsubscript{2}O} nanoparticles},
  Journal = {Phys. Rev. B},
  Year = {2021},
  Volume = {103},
  Number = {24},
  Month = {JUN 21},
  DOI = {10.1103/PhysRevB.103.245426},
  Article-Number = {245426},
  ISSN = {2469-9950},
  EISSN = {2469-9969}
}

@article{konzelmann2020,
  Author = {Konzelmann, Annika and Frank, Bettina and Giessen, Harald},
  Title = {Quantum confined {Rydberg} excitons in reduced dimensions},
  Journal = {J. Phys. B: At. Mol. Opt. Phys.},
  Year = {2020},
  Volume = {53},
  Number = {2},
  Month = {JAN 28},
  DOI = {10.1088/1361-6455/ab56a9},
  Article-Number = {024001},
  ISSN = {0953-4075},
  EISSN = {1361-6455}
}

@article{einevoll1992,
  Author = {Einevoll, G T},
  Title = {confinement of excitons in quantum dots},
  Journal = {Phys. Rev. B},
  Year = {1992},
  Volume = {45},
  Number = {7},
  Pages = {3410-3417},
  Month = {FEB 15},
  DOI = {10.1103/PhysRevB.45.3410},
  ISSN = {0163-1829}
}

@article{iotti1997,
  Author = {Iotti, R C and Andreani, L C},
  Title = {Crossover from strong to weak confinement for excitons in shallow or
    narrow quantum wells},
  Journal = {Phys. Rev. B},
  Year = {1997},
  Volume = {56},
  Number = {7},
  Pages = {3922-3932},
  Month = {AUG 15},
  DOI = {10.1103/PhysRevB.56.3922},
  ISSN = {2469-9950},
  EISSN = {2469-9969}
}

@article{david2022,
  Author = {David, Aurelien and Weisbuch, Claude},
  Title = {Excitons in a disordered medium: A numerical study in {InGaN} quantum
    wells},
  Journal = {Phys. Rev. Res.},
  Year = {2022},
  Volume = {4},
  Number = {4},
  Month = {OCT 4},
  DOI = {10.1103/PhysRevResearch.4.043004},
  Article-Number = {043004},
  EISSN = {2643-1564}
}

@article{que1992,
  Author = {Que, W M},
  Title = {Excitons in quantum dots with parabolic confinement},
  Journal = {Phys. Rev. B},
  Year = {1992},
  Volume = {45},
  Number = {19},
  Pages = {11036-11041},
  Month = {MAY 15},
  DOI = {10.1103/PhysRevB.45.11036},
  ISSN = {0163-1829}
}

@article{holtkemper2020,
  Author = {Holtkemper, M. and Quinteiro, G. F. and Reiter, D. E. and Kuhn, T.},
  Title = {Selection rules for the excitation of quantum dots by spatially
    structured light beams: Application to the reconstruction of higher
    excited exciton wave functions},
  Journal = {Phys. Rev. B},
  Year = {2020},
  Volume = {102},
  Number = {16},
  Month = {OCT 30},
  DOI = {10.1103/PhysRevB.102.165315},
  Article-Number = {165315},
  ISSN = {2469-9950},
  EISSN = {2469-9969}
}

@article{lloyd-hughes2012,
  author = {Lloyd-Hughes, J. and Jeon, T I},
  title = {A Review of the Terahertz Conductivity of Bulk and Nano-Materials},
  journal = {J Infrared Milli Terahz Waves},
  volume = {33},
  pages = {871–925},
  year = {2012},
  doi = {10.1007/s10762-012-9905-y}
}

@article{richter2017,
  title = {Nanoplatelets as material system between strong confinement and weak confinement},
  author = {Richter, Marten},
  journal = {Phys. Rev. Mater.},
  volume = {1},
  issue = {1},
  pages = {016001},
  numpages = {9},
  year = {2017},
  month = {Jun},
  publisher = {American Physical Society},
  doi = {10.1103/PhysRevMaterials.1.016001}
}

@article{raikh2021,
  title = {Size quantization of an exciton: A toy model of the “dead layer”},
  journal = {Solid State Communications},
  volume = {330},
  pages = {114265},
  year = {2021},
  issn = {0038-1098},
  doi = {10.1016/j.ssc.2021.114265},
  url = {https://www.sciencedirect.com/science/article/pii/S0038109821000740},
  author = {M.E. Raikh}
}

@article{fonoberov2004,
  title = {Radiative lifetime of excitons in {ZnO} nanocrystals: The dead-layer effect},
  author = {Fonoberov, Vladimir A. and Balandin, Alexander A.},
  journal = {Phys. Rev. B},
  volume = {70},
  issue = {19},
  pages = {195410},
  numpages = {5},
  year = {2004},
  month = {Nov},
  publisher = {American Physical Society},
  doi = {10.1103/PhysRevB.70.195410}
}

@misc{supplemental,
  note = "See Supplemental Material at http://link.aps.org/supplemental/10.1103/21qw-q87r
  for crossover to noninteracting e-h pairs in Si and justification of WCR approximation."
}


\clearpage
\onecolumngrid

\section*{Supplemental material\\Terahertz response of confined electron-hole pair: crossover between strong and weak confinement}%

\renewcommand{\thefigure}{S\arabic{figure}}
\renewcommand{\figurename}{FIG.}
\setcounter{section}{0}
\setcounter{figure}{0}

\section*{I. Crossover to non-interaction e-h pairs in indirect Si}

\begin{figure}[bh]
    \centerline{\includegraphics[width=11cm]{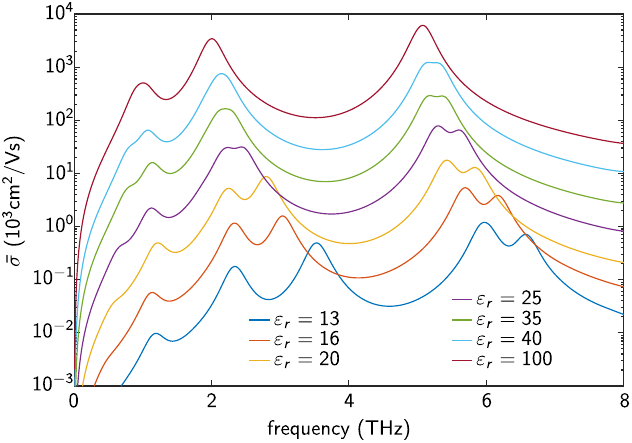}}
    \caption{Real part of normalized conductivity spectra of Si nanocrystal with $A=10$~nm
        with relative permittivity $\varepsilon_r$ changing between 100 (diminished electron-hole Coulomb
        interaction) and 13 (full interaction). The curves for $\varepsilon_r>13$ are
        vertically shifted.\label{fig:crossover}}
\end{figure}

The splitting of resonances in indirect Si in the strong confinement regime by the Coulomb interaction can 
be visualized by artificially tuning the strength of the electron-hole interaction by setting the 
relative permittivity $\varepsilon_r$ from a large value down to the actual value valid for Si. This crossover is 
plotted in Fig.~\ref{fig:crossover} and unambiguously shows that the lowest-energy resonance and
the high-frequency pair of resonance have origin in the electron coupling to radiation while
the middle two peaks are hole-like. These conclusions are in accordance with the analysis in Section
IV.B in the main text and Fig.~2(c).

\clearpage
\section*{II. Verification of the wavefunction separation}

\begin{figure}[bh]
    \centerline{\includegraphics[width=18cm]{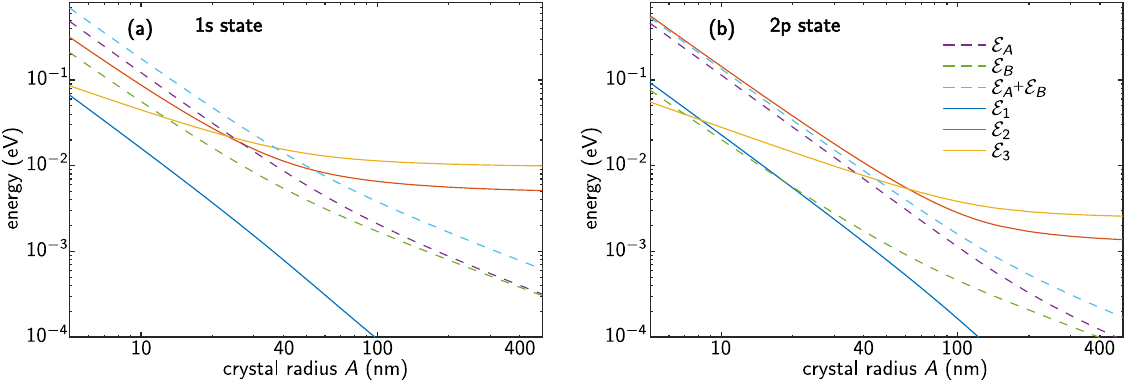}}
    \caption{Dependency of mean energies ${\cal E}_j$ using separated solutions (A2) for the
        $(1s)_w$ state (a) and $(2p)_w$ state (b) of the exciton. Neglected terms are plotted by dashed lines.\label{fig:energ}}
\end{figure}

Assumption of the separated wavefunction in Eq.~(A2) and further consideration of $r$ being a parameter
of the center-of-mass (COM) part $\Psi(\vec{R};r)$ leads to solvable equations (A3)--(A4) but neglects
some terms of Eq.~(A1). Our aim is to justify this neglection, in other words to show that contribution
of the neglected terms to the total energy is only a minor correction in the weak confinement regime (WCR).
To this end, we evaluate the magnitude of the mean energy of the neglected terms and compare them to
the rest of the terms. Since the full solution $\Phi(\vec{R},\vec{r})$ of Eq.~(A1) is not known,
we take our solution of the form of Eq.~(A2), consider that the relative coordinate $r$ is the argument
of the function $\Psi(\vec{R},r)$ and evaluate the neglected contributions:
\begin{eqnarray*}
    {\cal E}_A&=&\frac{\hbar^2}{2\mu}\bigg|\int |\psi(\vec{r})|^2\Big[\Psi^*(\vec{R},r)\nabla_r^2\Psi(\vec{R},r)\Big]\bigg|\,,\\
    {\cal E}_B&=&\frac{\hbar^2}{\mu}\bigg|\int\Big[\psi^*(\vec{r})\nabla_r\psi(\vec{r})\Big]\Big[\Psi^*(\vec{R},r)\nabla_r\Psi(\vec{R},r)\Big]\bigg|\,.
\end{eqnarray*}
Those are compared to the terms which remain in Eqs.~(A3)--(A4):
\begin{eqnarray*}
    {\cal E}_1&=&\frac{\hbar^2}{2M}\bigg|\int |\psi(\vec{r})|^2\Big[\Psi^*(\vec{R},r)\nabla_R^2\Psi(\vec{R},r)\Big]\bigg|\,,\\
    {\cal E}_2&=&\frac{\hbar^2}{2\mu}\bigg|\int |\Psi(\vec{R},r)|^2\Big[\psi^*(\vec{r})\nabla_r^2\psi(\vec{r})\Big]\bigg|\,,\\
    {\cal E}_3&=&\frac{e^2}{4\pi\varepsilon}\bigg|\int |\Psi(\vec{R},r)|^2\Big[\psi^*(\vec{r})\frac{1}{r}\psi(\vec{r})\Big]\bigg|\,.
\end{eqnarray*}
The above integrations are performed over both variables $\vec{r}$ and $\vec{R}$. The resulting dependencies 
of ${\cal E}_j(A)$ are plotted in Fig.~\ref{fig:energ}. Clearly, the terms ${\cal E}_A$ and ${\cal E}_B$ are small
compared to the terms which arise in Eqs.~(A3)--(A4) in the WCR. Note that the transition
from the WCR to the intermediate confinement regime (ICR) is defined by the crystal radius $A<10a_X$ 
where $a_X=15$~nm is the exciton Bohr radius in GaAs and even beyond this transition radius, the
two neglected terms are of minor importance. This finding justifies the separation of the wavefunction
in Eq.~(A2) in the WCR. In the ICR, the solution suffers from missing correlations between the COM and internal
degrees of freedom, the consequences of which are discussed in the main article. Also, the energies of the resulting states
become shifted compared to the full solution, and the upper limit of its error can be estimated as ${\cal E}_A+{\cal E}_B$. Nevertheless, the WCR solution reveals that the terms included in the solution after separation
are dominant over the neglected terms down to the crystal size $A\approx50$~nm, thus the WCR can be used as an estimate solution in the upper half of the ICR (large crystals).

\end{document}